\documentclass[11pt]{article}
\usepackage{amssymb}
\usepackage{gensymb}
\usepackage{amsmath}
\usepackage{amsthm}
\usepackage{textcomp}
\usepackage{hyperref}
\usepackage{afterpage}
\usepackage[margin=0.9in]{geometry} % Adjust margin as needed
\usepackage{comment}
\usepackage{graphicx} % for placing figures
\usepackage{subcaption}
\usepackage{float} % For allowing graphics, and other objects, to float around the page
\usepackage{url}
\usepackage{color} %or {xcolor}. % Allow text to be colored
\usepackage{enumitem} % for enumerating items
\usepackage{adjustbox}
\usepackage{titlesec}
\usepackage{stfloats}
\usepackage{setspace}
\usepackage{multirow} % multiple rows in tables
\usepackage{siunitx} % SI units for consistency
\usepackage[version=4]{mhchem} % chemical equations
\usepackage[capitalise, nameinlink, noabbrev]{cleveref}
\usepackage{soul}
\usepackage{tabularx}
\usepackage{longtable}
\usepackage{array}
\usepackage{booktabs}
\usepackage{geometry}
\usepackage{lscape} % Optional if using landscape for wide tables
\usepackage{caption}
\usepackage{multirow}
\usepackage{tabularx}
\usepackage{ragged2e}
\usepackage{lmodern}
\usepackage{authblk}
\usepackage[backend=biber,style=numeric,sorting=none]{biblatex}
\hypersetup{
    colorlinks=true,
    linkcolor=blue,
    filecolor=green,      
    urlcolor=blue,
    citecolor=blue
    } 
\newcolumntype{Y}{>{\RaggedRight\arraybackslash}X}
\title{\bf Synthesis and Characterization of Ultrasonically Atomized Al-Based Alloy Powders for Tunable Thermal Reactivity}

\author[1]{Chetan Singh\thanks{Corresponding author: \texttt{csingh9@jh.edu}}}
\author[2]{Ava Goglia}
\author[2]{Peter Mastracco}
\author[1,3]{Michael Flickinger}
\author[1,3]{Laszlo J.\ Kecskes}
\author[2]{Paulette Clancy}
\author[1,3,4]{Timothy P.\ Weihs}

\affil[1]{Hopkins Extreme Materials Institute, Johns Hopkins University, Baltimore, MD 21218, USA}
\affil[2]{Department of Chemical and Biomolecular Engineering, Johns Hopkins University, Baltimore, MD 21218, USA}
\affil[3]{Department of Materials Science and Engineering, Johns Hopkins University, Baltimore, MD 21218, USA}
\affil[4]{Department of Mechanical Engineering, Johns Hopkins University, Baltimore, MD 21218, USA}

\date{}

\begin{document}
\maketitle
\vspace{-2em}
\begin{abstract}
\normalsize
Reactive aluminum (Al) alloy powders are promising for advanced manufacturing, joining, and energetic applications, yet scalable routes that couple \emph{controlled} reactivity with safe handling remain limited. While nanoscale Al powders ignite readily, their agglomeration, handling, and safety limit broad deployment. Here, we manufacture micron-sized Al-based powders produced by ultrasonic atomization (UA), targeting a balance of enhanced reactivity and process robustness. Binary systems (Al--Cu, Al--Si, Al--Mg) and pure Al were synthesized, and their morphology, phases present, thermal stability, and oxidation behavior were characterized using XRD, SEM, and DTA/TGA in an Ar/\ce{O2} environment. We show that alloy selection and UA-controlled microstructure can modify the native \ce{Al2O3} passivation, alter oxidation pathways, and shift thermal onsets/exotherms. The results establish a manufacturing-forward framework for designing micron-sized powders with tunable ignition/oxidation behavior. 
\end{abstract}


\section{Introduction}
Reactive aluminum (Al) powders are central to energetics, joining/brazing, and additive manufacturing workflows that benefit from controllable heat release and ignition sensitivity \cite{price2000combustion_Al, yen2012reactive}. Nanoscale powders offer low ignition temperatures but suffer from aggregation, moisture uptake, and safety constraints, complicating metering and large-scale handling \cite{dreizin2009metal, wainwright2022commercial_Al, dreizin2000phase_combustion}. In contrast, micron-sized powders exhibit excellent flowability and dosing fidelity, but their thermal response is strongly inhibited by the native \ce{Al2O3} passivation layer \cite{sundaram2016general_nano_micron_Al, dreizin2015correlating_Al, nie2016initial_Dreizin_Al}. As a result, process routes that manufacture micron powders with tunable reactivity via controllable microstructures and surface architectures are of central interest to both equipment suppliers and end users in additive manufacturing and energetics related sectors.

For spherical micron-scale powders heated in oxidative environments, ignition typically proceeds through: (i) preheating of the metal core under a diffusion-limiting amorphous alumina; (ii) oxygen transport through and/or around the oxide; (iii) structural evolution of the oxide (densification, transformation to $\gamma$ and $\alpha$-Al2O3); and (iv) a rapid exothermic regime once metal/oxygen contact becomes sufficiently direct \cite{dreizin2015correlating_Al, hasani2012oxidation_Al}. Consequently, two practical materials strategies can enable reactivity-on-demand in micron-scale Al-based powders: (i) introduce near-surface phases that deliver an early exotherm or catalyze oxygen transport, and/or (ii) disrupt the continuity or protective character of the \ce{Al2O3} scale to accelerate oxygen ingress. These strategies can help us tune the reactivity of micron powders while retaining the logistical and safety benefits of micron size. 

Alloy-chemistries offer pathways to tune reactivity. Multiple alloying approaches are known to modify oxidation/ignition behavior at the particle scale: (i) Interfacial exotherm motifs, wherein near-surface phases (e.g. \ce{CuO}) react energetically with Al and locally preheat the particle \cite{matsuda2017multiple_Al}; (ii) Passivation-disruption motifs, wherein alloying promotes stress, porosity, or alternative oxides (e.g., \ce{MgO}/\ce{MgAl2O4}) that degrade the protective alumina \cite{trunov2005ignition_Al}; and (iii) Low-melting/volatile constituents (e.g. Ga, Bi) that redistribute early or open transport pathways during heating \cite{gandhi2024effect_Al_Ga}. The practical challenge is not just identifying chemistries, but manufacturing them reproducibly as flowable micropowders with controlled microstructure at industrially relevant yields.

Ultrasonic atomization (UA) is an attractive manufacturing method for both lab-scale alloy development and scalable powder production. It offers clean processing under inert gas, which helps minimize oxidation and contamination during melting \cite{priyadarshi2024new_ultrasonic, caccioppoli2002fabrication_ultrasonic}. UA also enables composition-sensitive solidification, producing spherical powders with microstructures—such as eutectic/lamellar, dendritic, or segregated skins—that can be tailored to influence oxidation behavior. Because the droplet thermal history depends on frequency, amplitude, melt superheat, and atmosphere, UA also offers a well-defined process window for suppliers to document and reproduce. This study reports an ultrasonic atomization process to fabricate pure Al and Al--Cu, Al--Si, and Al--Mg powders, followed by structural (SEM, XRD) and thermal (DTA/TGA) characterization. These compositions were selected because Cu, Si, and Mg each influence oxidation in different ways, offering a useful contrast in behavior. By comparing these systems, we aim to understand how alloying modifies the conventional ``Al core--\ce{Al2O3} shell'' architecture and how these changes affect the onset of oxidation and heat flow in an Ar/\ce{O2} environment.

\section{Materials and Methods}

\subsection{Ultrasonic atomization}
Aluminum alloy powders were synthesized from high-purity elemental feedstocks (ThermoFisher Scientific) using an ultrasonic atomization system (Amazemet). The elemental mixture was placed in a boron nitride-coated graphite crucible housed within an induction heating chamber and heated stepwise to \SI{200}{\degreeCelsius}, \SI{500}{\degreeCelsius}, and finally \SI{800}{\degreeCelsius}. At each temperature stage, the chamber was purged with argon to maintain an inert atmosphere and remove any adsorbed species. The alloy melt was held at \SI{800}{\degreeCelsius} for 5 minutes to ensure homogeneity.
A pressure differential of approximately \SI{-50}{\milli\bar} was initially established between the induction chamber and the spray chamber containing the sonotrode. Upon removal of the sealing rod, the pressure in the induction chamber was gradually increased from \SI{150}{\milli\bar} to \SI{850}{\milli\bar} over a 50-second interval, enabling a controlled and continuous flow of molten metal onto the atomizing surface. The molten stream impinged upon a carbon fiber composite (CFC) plate attached to a \SI{60}{\kilo\hertz} sonotrode. In the initial moments, spattering was observed due to insufficient wetting of the CFC surface; however, stable atomization was quickly established once proper wetting was achieved. The typical atomization yield in the sub-\SI{300}{\micro\meter} size range was approximately 80–90\% by weight. Representative alloy compositions included pure Al (control), Al--33Cu (33\,wt.\% Cu), Al--13Si (13\,wt.\% Si), and Al--33Mg (33\,wt.\% Mg). These compositions were selected near their respective eutectic points to promote the formation of alternating lamellar microstructures. Since eutectics exhibit the lowest melting points in their systems, they enable exploration of different levels of superheating and their effects on the resulting microstructure in a follow-on study.

\subsection{Particle size analysis}
Particle size distributions (PSD) were measured using a Horiba LA-950 V2 laser diffraction particle size analyzer. The as-atomized powders were first sieved to remove coarse particles and flakes larger than 300 µm, which are commonly generated during the initial stages of ultrasonic atomization due to incomplete droplet breakup or wetting of the sonotrode surface. For each measurement, the powders were dispersed in isopropanol and ultrasonicated to break apart any agglomerates. The reported particle size values represent the average of six independent measurements to ensure reproducibility and statistical confidence.

\subsection{X-ray Diffraction (XRD)}
Powders were filled into a low-background silicon sample holder to minimize preferred orientation effects. X-ray diffraction was performed using a Malvern Panalytical Aeris diffractometer equipped with a PIXcel1D-Medipix3 detector and Cu K\(\alpha\) radiation (\(\lambda = 1.5406\)\,Å), operated at \SI{40}{\kilo\volt} and \SI{7.5}{\milli\ampere}. Scans were collected over a \(2\theta\) range of \(10^{\circ}\) to \(90^{\circ}\) with a step size of \(0.02^{\circ}\), and the intensity was summed over two scans to improve signal-to-noise ratio.
Post-processing of the diffraction data, including background subtraction, sample misalignment correction, K\(\alpha_2\) stripping, peak identification, phase matching, and profile fitting, was performed using HighScore Plus4 analysis software.

\subsection{SEM/EDS and Phase Analysis}
Powder morphology and near-surface microstructure were examined using a Tescan MIRA 3 GM field-emission scanning electron microscope (FE-SEM). Samples were mounted on aluminum stubs using carbon adhesive tape. Imaging was performed at an accelerating voltage of \SI{15}{\kilo\volt} using a backscattered electron detector to enhance atomic number contrast, with a working distance of 4–6\,mm.

\subsection{Thermal Analysis}

Thermal behavior of the powders was characterized using simultaneous thermogravimetric analysis and differential scanning calorimetry (TGA/DSC) on a TA Instruments Discovery SDT 650 system. Measurements were conducted under a flowing gas mixture of argon and oxygen (80/21 vol.\%) at a constant heating rate of \SI{20}{\celsius\per\minute} to \SI{1200}{\celsius}. Approximately 10–15\,mg of powder, sieved to the 32–53\,\si{\micro\meter} size range, was used per trial. Samples were placed in standard alumina crucibles. A minimum of three replicate measurements were performed for each alloy composition to ensure reproducibility of thermal profiles.

\section{Results and Discussion}

\subsection{Particle size distributions, and implications for atomization}
\begin{figure}[h!]
  \centering
  \includegraphics[height=7cm, width=\linewidth, keepaspectratio]{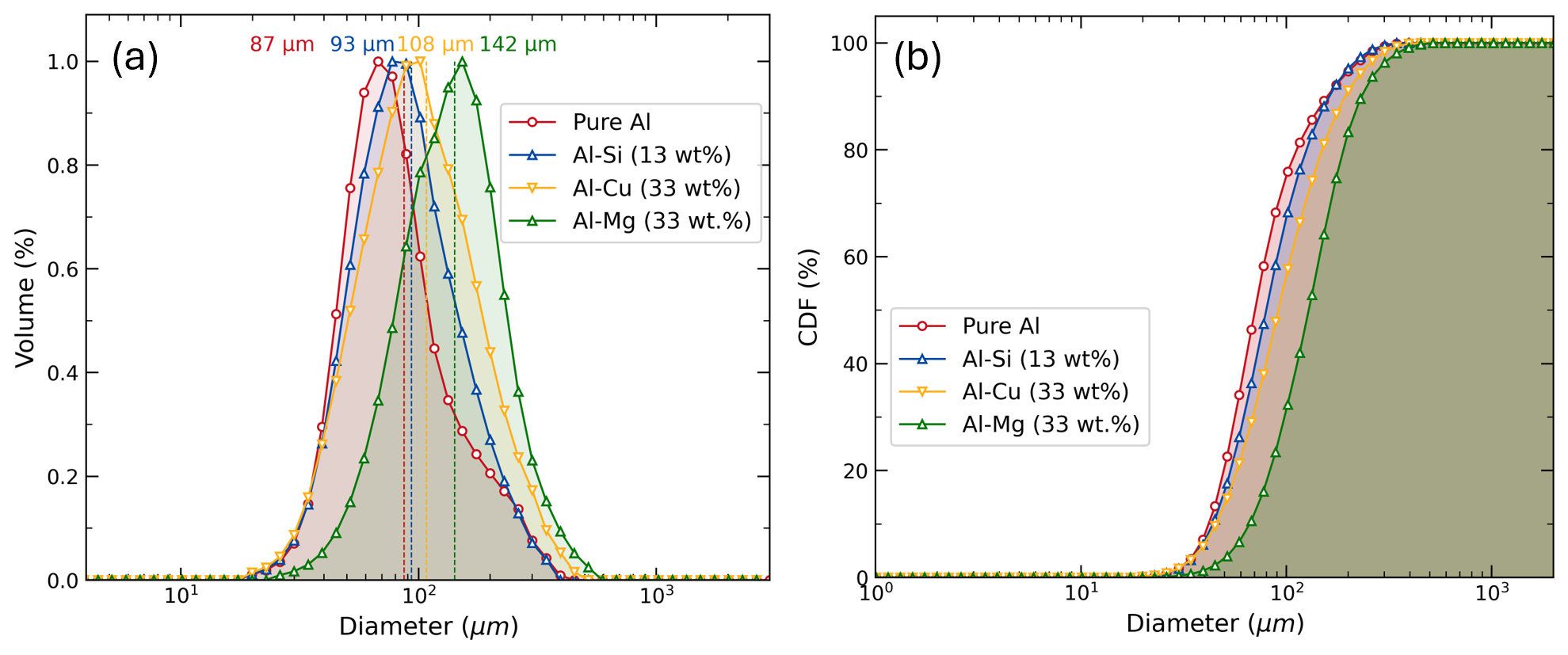}
  \caption{(a) Volume-based PSDs for as-atomized Al-alloy powders; (b) cumulative distribution functions (CDFs) of the volume-based PSDs.}
  \label{fig:psd}
\end{figure}

The normalized particle size distributions (PSDs) in \cref{fig:psd}a show clear alloy-dependent shifts in droplet breakup and solidification under ultrasonic atomization. Pure Al exhibits the narrowest PSD (mean $\sim$87\,\si{\micro\meter}), consistent with low melt viscosity/surface tension favoring efficient cavitation and capillary-wave disintegration \cite{priyadarshi2024new_ultrasonic}. Alloying broadens and coarsens the distribution: Al--Si (mean $\sim$93\,\si{\micro\meter}) and Al--Cu (mean $\sim$108\,\si{\micro\meter}) lie intermediate, while Al--Mg (33\,wt.\%) is broadest and coarsest (mean $\sim$142\,\si{\micro\meter}), plausibly reflecting higher effective viscosity. The cumulative distributions in \cref{fig:psd}b echo these trends (steeper CDFs for pure Al; more gradual CDFs for Al--Mg).
\Cref{tab:psd} summarizes the $D_{10}$, $D_{50}$, and $D_{90}$ values for the as-atomized powders (with flakes and particles larger than 300 µm removed). All structural and thermal characterizations discussed in subsequent sections were conducted using the 32–53 µm sieve fraction. While the PSD serves as a relevant process metric for manufacturing yield, the use of the 32–53 µm cut ensures a consistent basis for XRD, SEM, TGA, and DSC comparisons. This approach isolates the effects of alloying additions by eliminating the size-dependent variability in the measured physical and thermal properties.

\begin{table}[h!]
  \centering
  \caption{Particle size distribution metrics for the produced Al-alloy powders (as-atomized).}
  \label{tab:psd}
  \begin{tabular}{lccc}
    \toprule
    Alloy & $D_{10}$ (\si{\micro\meter}) & $D_{50}$ (\si{\micro\meter}) & $D_{90}$ (\si{\micro\meter}) \\
    \midrule
    Al        & 42  & 70  & 159 \\
    Al--Si    & 44  & 80  & 162 \\
    Al--Cu    & 45  & 91  & 194 \\
    Al--Mg    & 66  & 128 & 230 \\
    \bottomrule
  \end{tabular}
\end{table}

\subsection{Phase constitution and solidification pathways}
\begin{figure}[h!]
  \centering
  \includegraphics[height=7cm, width=\linewidth,keepaspectratio]{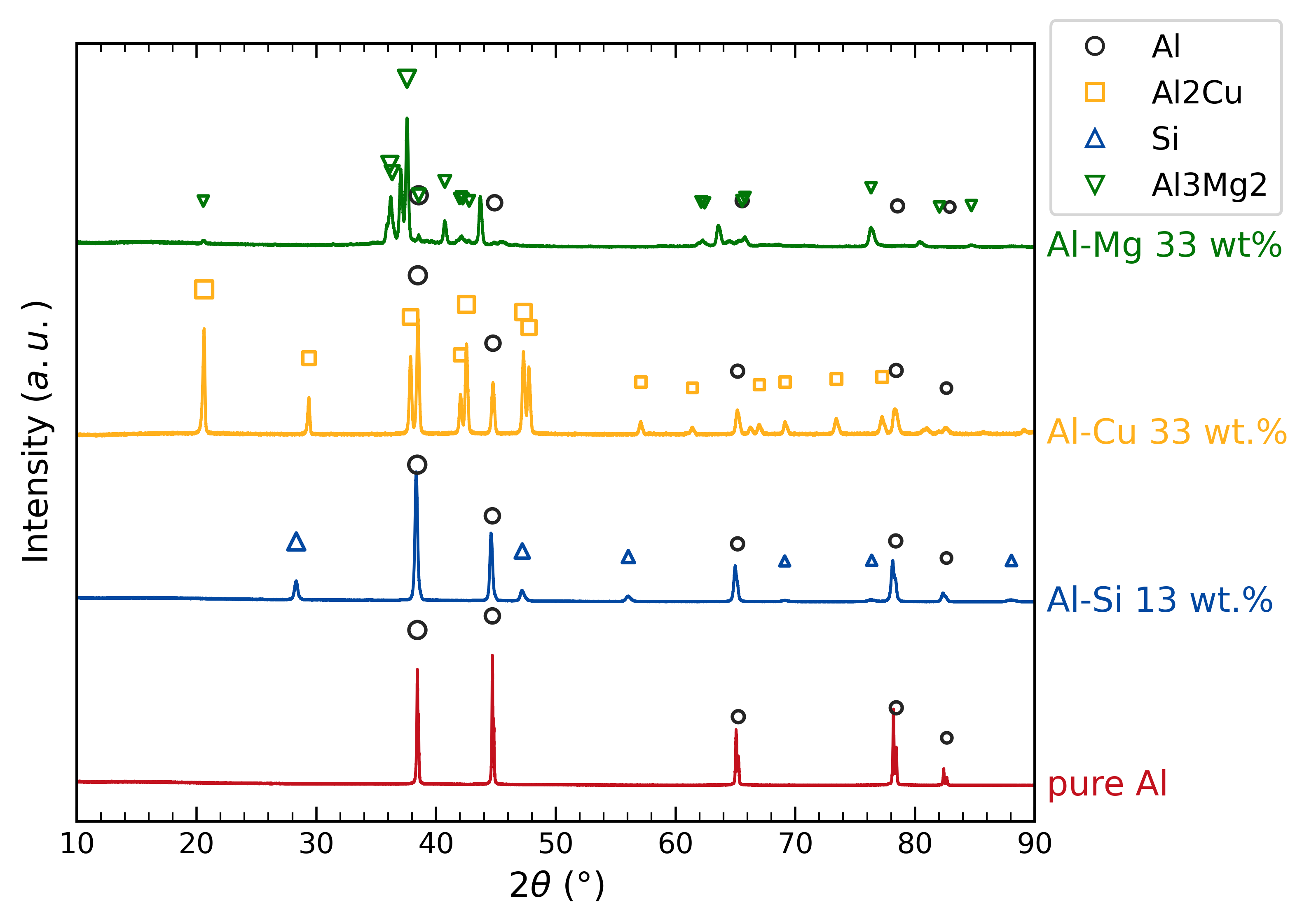}
  \caption{XRD patterns of as-atomized Al-alloy powders. Phase identification includes fcc-Al for all powders; Si (diamond cubic) in Al--Si; $\theta$-\ce{Al2Cu} in Al--Cu; and $\beta$-\ce{Al3Mg2} in Al--Mg.}
  \label{fig:xrd}
\end{figure}

\Cref{fig:xrd} overlays the powder XRD patterns. Pure Al shows fcc-Al (Fm$\bar{3}$m), as expected for single-phase droplets \cite{Massalski1990_Binary}. Al--13\,wt.\%\,Si exhibits fcc-Al and Si (diamond cubic), indicative of a eutectic/near-eutectic morphology \cite{Murray1984_AlSi}. Al--33\,wt.\%\,Cu shows fcc-Al plus $\theta$-\ce{Al2Cu} reflections, consistent with eutectic solidification and/or intermetallic precipitation during rapid cooling \cite{Zobac2019_AlCu}. In Al--33\,wt.\%\,Mg, fcc-Al coexists with $\beta$-\ce{Al3Mg2}, evidencing limited Mg solubility and secondary-phase formation \cite{Okamoto1998_AlMg, mukherjee2010formationAlMg_Al3Mg2}. These phases formed in as-atomized powders align with the observed surface textures and are useful predictors for oxidation as noted in the next subsection.

\subsection{Particle morphology and near-surface features}
\begin{figure}[t]
  \centering
  \includegraphics[height=7cm, width=\linewidth,keepaspectratio]{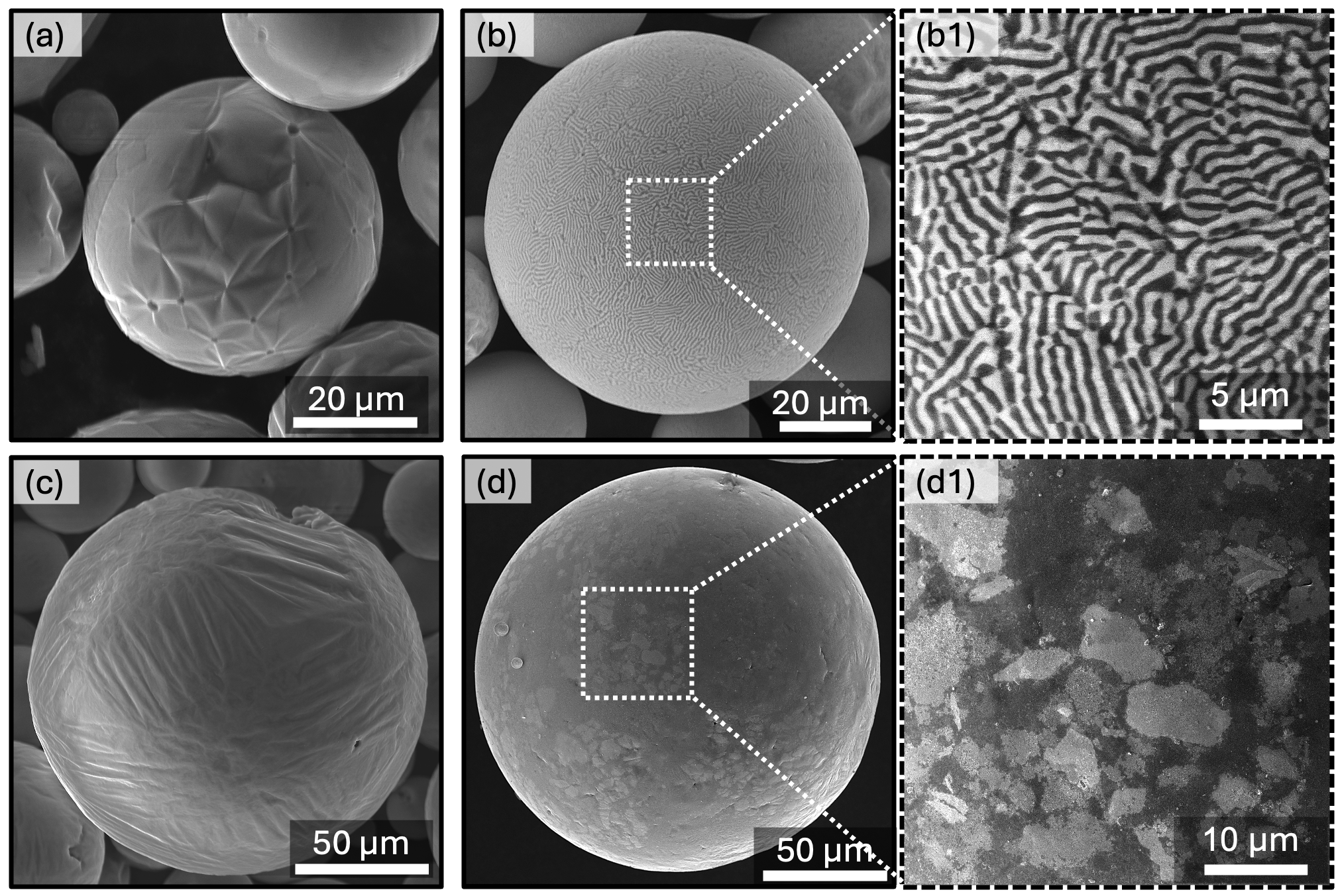}
  \caption{Representative SEM images of as-atomized powders: (a) Pure Al; (b) Al--33\,wt.\%\,Cu; (b1) higher-magnification lamellae; (c) Al--13\,wt.\%\,Si; (d) Al--33\,wt.\%\,Mg; (d1) higher magnification indicating likely $\beta$-\ce{Al3Mg2}/oxide contrast.}
  \label{fig:sem}
\end{figure}

Representative SEM micrographs (\cref{fig:sem}) reveal composition-sensitive droplet solidification and surface architectures. Pure Al (panel a) shows faceted surfaces and radial shrinkage wrinkles from early shell solidification followed by core contraction. Al--33\,wt.\%\,Cu (b,b1) presents a dense, uniform eutectic lamellar texture; the fine alternating Al/\ce{Al2Cu} lamellae inferred from the surface are consistent with the $\theta$-\ce{Al2Cu} reflections in XRD. Al--13\,wt.\%\,Si (c) exhibits pronounced creases/wrinkling, plausibly tied to Al/Si thermal-mismatch and eutectic front instabilities \cite{ikhmayies2019phaseAlSi}. Al--33\,wt.\%\,Mg (d,d1) appears smoother but mottled; the contrast variations likely reflect $\beta$-\ce{Al3Mg2} segregation and/or early oxidation (Mg-rich regions) given high oxygen affinity of Mg \cite{nie2016oxidationAlMg, wainwright2023comparing_Al_Ti_Zr_Mg}. These alloy-dependent topographies foreshadow distinct oxidation pathways: Cu-bearing eutectic near the surface can modify interfacial reactions \cite{Zobac2019_AlCu, Wang2019_Al2Cu_Revisited}; Si-rich microstructures maintain an Al-dominated oxide \cite{ikhmayies2019phaseAlSi}; Mg-rich features can promote preferential Mg oxidation \cite{nie2016oxidationAlMg}.

\subsection{Thermal stability and oxidation behavior}
\begin{figure}[h!]
  \centering
  \includegraphics[height=7cm, width=\linewidth,keepaspectratio]{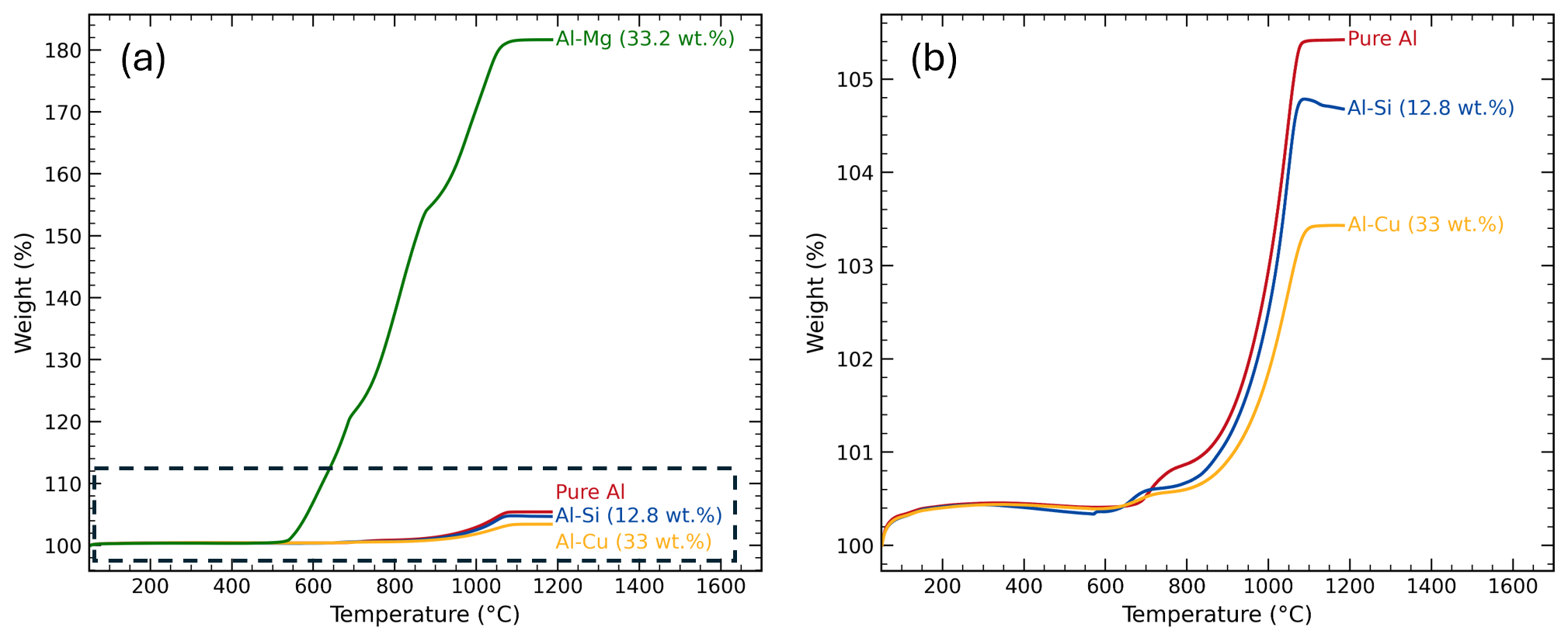}
  \caption{TGA curves in Ar/\ce{O2} (80/20\,vol.\%): (a) full-scale comparison; (b) zoom on low-gain alloys.}
  \label{fig:tga}
\end{figure}

TGA comparisons in Ar/\ce{O2} (80/20\,vol.\%) highlight a strong composition effect (\cref{fig:tga}). Al--Mg (33.2\,wt.\%) shows a large mass gain approaching $\sim$82\% by \SI{1100}{\celsius}, consistent with preferential Mg oxidation and the formation of \ce{MgO}/\ce{MgAl2O4}-rich phases that are non-protective under these conditions \cite{wu2019oxidation_AlMg, aly2015ignition_AlMg}. In contrast, pure Al, Al--Si, and Al--Cu gain only $\sim$3--6\% over the same range, with gradual onsets above $\sim$\SI{600}{\celsius}, as expected for alumina-controlled oxidation of micron Al \cite{dreizin2015correlating_Al}.

Zoomed TGA curves for the low-mass gain set (Al, Al--13\,wt.\%\,Si, Al--33\,wt.\%\,Cu) in \cref{fig:tga}b show similar slow mass gains initiating near \SI{650}{\celsius}, i.e., where protective \ce{Al2O3} becomes more transport-limiting \cite{dreizin2015correlating_Al}. The slightly lower gain observed for Al--Cu relative to Al/Al--Si is consistent with Cu-bearing intermetallics altering near-surface transport/chemistry and modestly impeding net oxidation in this temperature window \cite{wu2019oxidation_AlCu, zhao2023highAlCuSi}. Together with XRD/SEM, these data indicate: (i) Si additions mainly retain a protective Al-oxide response; (ii) Cu additions can reduce mass gain slightly (interfacial/kinetic effects); and (iii) Mg-rich compositions shift the system to a Mg‑dominated oxidation pathway with much larger mass uptake.

\begin{figure}[h!]
  \centering
  \includegraphics[height=7cm, width=\linewidth,keepaspectratio]{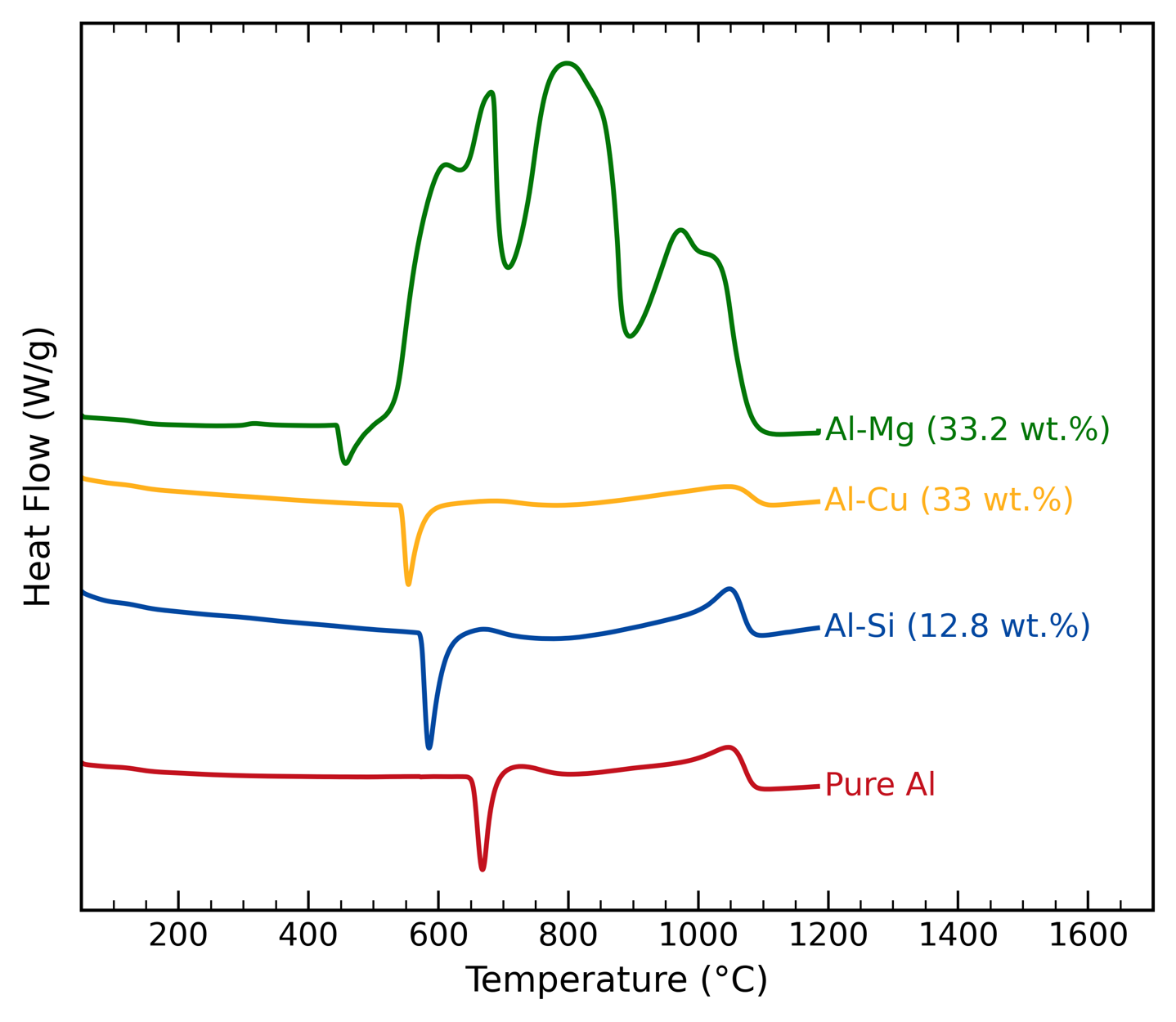}
  \caption{DSC heat-flow curves for the 32--53\,\si{\micro\meter} fraction under Ar/\ce{O2} (80/20\,vol.\%).}
  \label{fig:dsc}
\end{figure}

DSC traces (\cref{fig:dsc}) corroborate alloy-specific thermal events. Pure Al shows a sharp melting endotherm near \SI{660}{\celsius}, followed by weak exothermic signatures at higher $T$ associated with oxidation after melt. Al--13\,wt.\%\,Si exhibits its principal melting near the Al--Si eutectic regime near \SI{580}{\celsius}  (primary endotherm below pure Al, with a subtle shoulder consistent with oxidation) \cite{Murray1984_AlSi, zhao2023highAlCuSi}. Al--33\,wt.\%\,Cu shows a strong endotherm near the Al--Cu eutectic domain ($\sim$\SI{550}{\celsius}) \cite{wu2019oxidation_AlCu, kakitani2019thermalAlCu}. Al--33.2\,wt.\%\,Mg displays a primary melting endotherm near \SI{450}{\celsius} followed by multiple broad features from $\sim$\SI{550}{\celsius} to beyond \SI{1100}{\celsius}, reflecting overlapping transformations/oxidation linked to $\beta$-\ce{Al3Mg2} and Mg-rich oxidation pathways \cite{Okamoto1998_AlMg, nie2016oxidationAlMg, wu2019oxidation_AlMg}. The DSC/TGA agreement with XRD/SEM strengthens the process–structure–reactivity link: the as-atomized phase/microstructure sets the accessible oxidation routes and their kinetics under identical test conditions.

\subsection{Implications for reactive applications and manufacturing quality control}
From a reactivity standpoint, the data suggest distinct application niches. Pure Al provides low oxidation and tight PSDs—well-suited where stability and flowability dominate. Al--Si introduces modest changes (surface complexity with largely Al-controlled oxidation). Al--Cu offers fine eutectic textures and slightly reduced oxidation, potentially useful where a small shift in ignition/oxidation threshold is advantageous, and the nanoscale lamellar architecture can be leveraged for further processing. Al--Mg maximizes potential heat release, evident by much greater oxidation; storage and handling of these powders will need tighter controls.

For manufacturing, lot-level acceptance metrics flow naturally from these results: (i) PSD targets and yield for the 32--53\,\si{\micro\meter} fraction; (ii) XRD phase confirmation (Al plus intended second phases; absence of extraneous oxides); and (iii) TGA/DSC windows (e.g., onset $T_{\mathrm{on}}$ and specific exotherm bounds for a standard Ar/\ce{O2} ramp). Such quality control criteria anchor reproducible production of reactivity-tuned micropowders.

\subsection{Future work}
This study establishes links between alloy composition and the morphological, thermal, and phase characteristics of ultrasonically atomized Al-based powders. Building on these findings, several ongoing efforts aim to extend the scope and impact of this work. First, reactivity assessments are currently limited to thermal analysis under oxidative conditions; wire-ignition and combustion testing are underway to provide a more direct evaluation of ignition sensitivity, combustion efficiency, and relevance to application-specific thresholds. Second, surface engineering strategies—such as low-temperature oxidative conditioning—and process optimization are being explored to tailor near-surface phases and morphologies, which are expected to play a critical role in controlling ignition and combustion behavior. Finally, the integration of CALPHAD-based thermodynamic modeling and machine learning approaches is planned as sufficient lot-wise and compositional data accumulate. These efforts aim to establish predictive relationships and guide future alloy design for scalable, high-performance reactive powders.

\section{Conclusion}

This study demonstrates the synthesis and characterization of reactive Al-based alloy powders produced via ultrasonic atomization, focusing on four representative compositions: pure Al, Al–Si (13\,wt.\%), Al–Cu (33\,wt.\%), and Al–Mg (33\,wt.\%). Particle size analysis revealed alloy-dependent shifts in droplet breakup behavior, with mean particle sizes ranging from 87\,µm (pure Al) to 142\,µm (Al–Mg), highlighting the role of melt viscosity and surface tension in determining atomization dynamics. X-ray diffraction confirmed the presence of primary fcc-Al in all compositions, along with distinct secondary phases such as $\beta$-Si, Al\textsubscript{2}Cu, and $\beta$-Al\textsubscript{3}Mg\textsubscript{2} depending on the alloying elements. Thermal analysis revealed enhanced reactivity in the Al–Mg system, evidenced by significant mass gain and multiple exothermic events at lower temperatures, whereas Al–Cu and Al–Si showed more modest oxidation behavior. SEM imaging further corroborated these findings, showing distinct surface morphologies and solidification features for each alloy system.

Collectively, the results highlight the effectiveness of ultrasonic atomization for producing compositionally tunable, micron-scale reactive powders with controllable size, morphology, and thermal behavior. These powders offer potential in a range of applications, from additive manufacturing to energetic materials and joining technologies. The findings also provide a foundation for future optimization of reactivity through surface engineering, alloy design, and predictive modeling approaches.

\section*{Acknowledgments}
This research was supported by the Department of Defense, Defense Threat Reduction Agency (DTRA) under the MSEE URA (Grant HDTRA1-20-2-0001). Chetan Singh, Michael R. Flickinger, Peter Mastracco, and Timothy P. Weihs received support through the MSEE URA. The authors gratefully acknowledge the Hopkins Extreme Materials Institute and the Department of Materials Science and Engineering at Johns Hopkins University for providing access to characterization and sample preparation facilities. 
The authors also thank Brennan McGrann and Grecia Muñoz González for their assistance with preliminary XRD and DTA experiments.

\section*{Data Availability}
Data (SEM images, XRD patterns, TGA/DSC curves, PSD files) will be provided upon reasonable request.

\printbibliography

\end{document}